\documentstyle[eqsecnum,aps]{revtex}

\begin{document}
\draft
\title{Gaussian Wave-functional Approach in \\
Thermofield Dynamics}
\author{Wen-Fa Lu}
\address{CCAST(World Laboratory) P.O. Box 8730, Beijing, 100080,
  \\ and \\
Physics of Department, Shanghai Jiao Tong University,
Shanghai 200030, China}
\date{\today}
\maketitle

\begin{abstract}
The Gaussian wavefunctional approach is developed in thermofield dynamics. We
manufacture thermal vacuum wavefunctional, its creation as well as annihilation
operators, and accordingly thermo-particle excited states. For a
($D+1$)-dimensional scalar field system with an arbitrary potential whose
Fourier representation exists in a sense of tempered distributions, we
calculate the finite temperature Gaussian effective potential (FTGEP), one- and
two-thermo-particle-state energies. The zero-temperature limit of each of them
is just the corresponding result in quantum field theory, and the FTGEP can
lead to the same one of each of some concrete models as calculated by the
imaginary time Green function.
\end{abstract}

The Gaussian wave-functional approach ( GWFA ) \cite{1} has become an important
non-perturbative tool of quantum field theory and condensed matter physics
\cite{2,3} since Stevenson's advocation in 1985 \cite{4}. It is simple,
feasible, effective, and can reveal useful non-perturbative information (at
least qualitatively). A result from the GWFA can give a helpful guidance or
be taken as a good start for a further investigation \cite{5}. Moreover, this
approach is often used for some novel investigations or new ideas \cite{6}.
So the GWFA was inevitably generalized to finite temperature field theory, by
a direct calculation of partition function from the first principle or by
calculating the real-time or imaginary-time Green functions \cite{7}. However,
thermofield dynamics \cite{8}, the third formalism of finite temperature field
theory, has its own advantages over the other two formalisms : the closed time
path and the imaginary time formalisms \cite{9}. It can reply to some questions
which are not possible within the framework of the closed time path formalism,
such as the structure of the thermal vacuum, the nature of the Goldstone states
, and so on \cite{9}. Also, Thermofield dynamics can describe very naturally
the time development for quantities near equilibrium \cite{9}. Hence it is
worth while generalizing the GWFA to thermofield dynamics. In 1986, through
considering the $\lambda\phi^4_{(0+1)}$ theory, I. Roditi gave the
quantum-mechanical finite temperature Gaussian effective potential ( FTGEP )
within the framework of thermofield dynamics \cite{10}. Recently, A. Mishra and
H. Mishra calculated the FTGEP of the $\lambda\phi^4_{(3+1)}$ theory by the
combination of the zero-temperature Bogoliubov and the thermal Bogoliubov
transformation techniques ( TDBTT ) \cite{11}. In this letter, we intend to
develop Roditi's pioneering work and establish the GWFA of thermofield
dynamics ( TDGWFA ) in $(D+1)$ dimensions.

We consider a relatively general model with the Lagrangian density
\begin{equation}
{\cal L}={\frac {1}{2}}\partial_\mu \phi_x \partial^\mu \phi_x
-V(\phi_x) \;,
\end{equation}
where $\phi_x\equiv \phi(\vec{x})$, and the potential $V(\phi_x)$ has a Fourier
reprensentation in a sense of tempered distributions \cite{12}. Potentials of
many models, such as various polynomial models, sine-Gordon and sinh-Gordon
models, have this property. We suppose that the field system Eq.(1) is immerged
into a thermal reservoir with a fixed temperature $T={\frac {1}{k_b\beta}}$ (
$k_b$ represents the Boltzmann constant), $i.e.$, we shall consider canonical
ensemble. ( As for the Grand canonical ensemble, the generalization is straight
foreward.) After manufacturing the thermal vacuum wavefunctional and the
thermal particle creation as well as annihilation operators, we shall calculate
the FTGEP, one- and two-thermo-particle energies, and then conclude this letter
by a brief discussion.

By the way, thermofield dynamics and the GWFA in quantum field theory will be
not introduced in this letter, and good exposition of them can be found in
Refs.~\cite{8,9} and refs.~\cite{1,2,3}, respectively.

Let us begin with quantum field theory. In the fixed-time functional
Schr\"odinger picture, the canonical conjugate operater to $\phi_x$ is $\Pi_x
\equiv -i{\frac {\delta}{\delta \phi_x}}$ with the commutator $[\phi_x,\Pi_y]
=i\delta(x-y)$. For the convenience of renormalizing the theory of the
system Eq.(1) later, we shall use the following normal-ordered Hamiltonian
density with respect to any normal-ordering mass $Q$ \cite{13}
\begin{equation}
{\cal N}_Q [{\cal H}_x]={\frac {1}{2}}\Pi^{2}_x
   +{\frac {1}{2}}(\nabla\phi_x)^2 -{\frac {1}{2}}I_0[Q^2]
   +{\frac {1}{4}}Q^2 I_1[Q^2] +{\cal N}_Q[V(\phi_x)]  \;\;,
\end{equation}
where $\nabla$ represents the gradient operator in $D$-dimensional
$\vec{x}$-space, $p=|\vec{p}|$, and $I_n[Q^2]=\int {\frac {d^D\vec{p}}
{(2\pi)^D}}{\frac {\sqrt{p^2+Q^2}}{(p^2+Q^2)^n}}$. As for
${\cal N}_Q[V(\phi_x)]$, making the Fourier transformation ( at least in a
sense of tempered distributions \cite{12} ) and using the Baker-Hausdorff
formula, one can have
\begin{equation}
{\cal N}_Q[V(\phi_x)] = \int^\infty_{-\infty}
 {\frac {d\Omega}{\sqrt{2\pi}}} {\tilde{V}}(\Omega)
       e^{i\Omega\phi_x}e^{{\frac {1}{4}}\Omega^2 I_1 [Q^2]} \;.
\end{equation}

Similar to the GWFA in quantum field theory \cite{1,3}, we take following
Gaussian wavefunctional as the quantum vacuum ansatz of $T=0$ field system
Eq.(1)
\begin{equation}
|\varphi> = N_fexp\{ -{\frac {1}{2}}\int_{x,y}
            (\phi_x - \varphi_x)f_{xy}(\phi_y - \varphi_y)\} \;,
\end{equation}
where $\varphi_x$ and $f_{xy}$ are variational parameter functions. $N_f$ is
some normalization constant, and depends upon $f_{xy}$. Due to the
translational invariance of vacuum, it is required that $f_{xy}=f(x-y)$. Also,
in view of the symmetric form of the vacuum, we assume that $f_{xy}=f_{yx}$ .
Besides, the inverse $f_{xy}^{-1}$ of $f_{xy}$ has to exist. It is easy to show
that $\varphi_x$ is the expectation value of the field operator with respect to
the above trial vacuum $|\varphi>$. The creation and the annihilation operators
to the vacuum Eq.(4) can be constructed as \cite{1,3}
\begin{equation}
A_f(\vec{p}) = ({\frac {1}{2(2\pi)^D f(\vec{p})}})^{1/2}\int_x
  e^{-i\vec{p}\cdot\vec{x}}[f(\vec{p})(\phi_x - \varphi_x)
+i\Pi_x]
\end{equation}
and
\begin{equation}
A^\dagger_f(\vec{p}) = ({\frac {1}{2(2\pi)^D f(\vec{p})}})^{1/2}\int_x
e^{i\vec{p}\cdot\vec{x}}
[f(\vec{p})(\phi_x - \varphi_x) - i\Pi_x]
\end{equation}
respectively, where $f(\vec{p})$ is the Fourier transform of $f_{xy}$. Of
course, one can check the relations $[A_f(\vec{p}),A^\dagger_f(\vec{p}')]=
\delta (\vec{p}'-\vec{p})$ and $A_f(\vec{p})|\varphi>=0$.

In order to calculate average value of any physical quantity, thermofield
dynamics needs the tilde system, the copy of the $T=0$ field system.
Corresponding to the above, the tilde system has the normal-ordered Hamiltonian
density
\begin{equation}
{\cal N}_Q[\tilde{\cal H}_x]={\frac {1}{2}}\tilde{\Pi}^{2}_x
   +{\frac {1}{2}}(\nabla\tilde{\phi}_x)^2 -{\frac {1}{2}}I_0[Q^2]
   +{\frac {1}{4}}Q^2 I_1[Q^2] +{\cal N}_Q[V(\tilde{\phi}_x)]
\end{equation}
and the quantum vacuum
\begin{equation}
|\tilde{\varphi}> = N_fexp\{-{\frac {1}{2}}\int_{x,y}
            (\tilde{\phi}_x - \varphi_x)f_{xy}(\tilde{\phi}_y - \varphi_y)\} \;,
\end{equation}
where $\tilde{\Pi}_x\equiv i{\frac {\delta}{\delta \tilde{\phi}_x}}$ conjugates
to the tilde field operator $\tilde{\phi}_x$ with the commutator
$[\tilde{\phi}_x,\tilde{\Pi}_y]=-i\delta(x-y)$. The appearance of the
additional minus in the commutation relation is due to the rule of the tilde
operation \cite{8,9}. Note that it is not necessary to distinguish the ordinary
real function and average values of the $T=0$ system with those of its tilde
system. From the rule of the tilde operation, the annihilation and the creation
operators to the tilde vacuum Eq.(8) read
\begin{equation}
\tilde{A}_f(\vec{p}) = ({\frac {1}{2(2\pi)^D f(\vec{p})}})^{1/2}\int_x
  e^{i\vec{p}\cdot\vec{x}}[f(\vec{p})(\tilde{\phi}_x - \varphi_x)
-i\tilde{\Pi}_x]
\end{equation}
and
\begin{equation}
\tilde{A}^\dagger_f(\vec{p}) = ({\frac {1}{2(2\pi)^D f(\vec{p})}})^{1/2}\int_x
    e^{-i\vec{p}\cdot\vec{x}}[f(\vec{p})(\tilde{\phi}_x - \varphi_x)
   +i\tilde{\Pi}_x]\;.
\end{equation}
Of course, the relations $[\tilde{A}_f(\vec{p}),\tilde{A}^\dagger_f(\vec{p}')]
 =\delta (\vec{p}'-\vec{p})$ and $\tilde{A}_f(\vec{p})|\tilde{\varphi}>=0$ hold.
Additionally, each operator of the tilde system commutes with that of the
original system.

Basing on the $T=0$ field system and its tilde partner, now we take the
following wavefunctional as the thermal vacuum ansatz of the $T\not= 0$ field
system
\begin{equation}
|\varphi,\beta>=exp\{i\int_{x,y}[(\phi_x -\varphi_x)g_{xy}(\beta)\tilde{\Pi}_y
         -\Pi_x g_{xy}(\beta)(\tilde{\phi}_y - \varphi_y)]\}
         |\varphi\tilde {\varphi}>  \;,
\end{equation}
where $|\varphi\tilde {\varphi}>$ represents the direct product of $|\varphi>$
and $|\tilde {\varphi}>$, $i.e.$,
\begin{equation}
|\varphi\tilde {\varphi}> = N^2_f exp\{-{\frac {1}{2}}\int_{x,y}[
           (\phi_x - \varphi_y)f_{xy}
         (\phi_y - \varphi_y)+(\tilde{\phi}_x - \varphi_x)f_{xy}
         (\tilde{\phi}_y - \varphi_y)]\} \;.
\end{equation}
Analogous with $f_{xy}$, $g_{xy}(\beta)=g_{yx}(\beta)=g(x-y,\beta)$, and it
is also a temperature-dependent variational parameter function to be
determined. When it acts on the thermal vacuum wavefunctional Eq.(11), any
operator of the original system works only for quantities of the original
system itself. So does that of the tilde system. According to Ref.~\cite{14},
one can have ( here the Baker-Hausdorff formula is invalid )
\begin{eqnarray}
&&exp\{i\int_{x,y}[(\phi_x -\varphi_x)g_{xy}(\beta)\tilde{\Pi}_y
         -\Pi_x g_{xy}(\beta)(\tilde{\phi}_y - \varphi_y)]\} \nonumber \\
&=&exp\{-\int '_{xpy}th(g(\vec{p},\beta))\Pi_x(\tilde{\phi}_y - \varphi_y)\}
  \nonumber \\  &&\cdot exp\{\int '_{xpy} ln(ch(g(\vec{p},\beta)))
         [(\phi_x -\varphi_x)\Pi_y
         +(\tilde{\phi}_y - \varphi_x)\tilde{\Pi}_x]\} \nonumber \\
&&  \cdot exp\{\int '_{xpy} th(g(\vec{p},\beta))(\phi_x -\varphi_x)\tilde{\Pi}_y
         \}  \;, \nonumber
\end{eqnarray}
where $\int '_{xpy} \equiv i\int_{x,y}\int
      {\frac {d^D\vec{p} e^{i\vec{p}\cdot (\vec{x}-\vec{y})}}{(2\pi)^D}}$,
      and $g(p,\beta)$ is the Fourier transform of the function $g_{xy}(\beta)$
. Hence one can write Eq.(11) as
\begin{eqnarray}
|\varphi,\beta>&=&N^2_f exp\{-{\frac {i}{2}}\int '_{xpy}[
                 (\phi_x ch(g(\vec{p},\beta))
                 -\tilde{\phi}_x sh(g(\vec{p},\beta))
                 - \varphi_x (1-sh(g(\vec{p},\beta))))   \nonumber  \\
          &\;\;& \cdot f(\vec{p})(\phi_y ch(g(\vec{p},\beta))
                 -\tilde{\phi}_y sh(g(\vec{p},\beta))
                 - \varphi_y (1-sh(g(\vec{p},\beta)))) \nonumber \\
         &\;\;& + (\tilde{\phi}_x ch(g(\vec{p},\beta))
                 -\phi_x sh(g(\vec{p},\beta))
                 - \varphi_x (1-th(g(\vec{p},\beta))
                              +sh(g(\vec{p},\beta))th(g(\vec{p},\beta))))
                 \nonumber \\
         &\;\;& \cdot f(\vec{p})(\tilde{\phi}_y ch(g(\vec{p},\beta))
                 -\phi_y sh(g(\vec{p},\beta))
                 - \varphi_y (1-th(g(\vec{p},\beta))
                              +sh(g(\vec{p},\beta))th(g(\vec{p},\beta)))) ]\}
                              \;\; \;.
\end{eqnarray}
This wavefunctional is just Gaussian for $\phi$ or for $\tilde{\phi}$.

For the thermal vacuum Eq.(11) or (13), one can manufacture the following
annihilation and creation operators:
\begin{eqnarray}
A_f(\vec{p},\beta) &=& ({\frac {1}{2(2\pi)^D f(\vec{p})}})^{1/2}\int_x
   \{f(\vec{p})[ e^{-i\vec{p}\cdot\vec{x}}ch(g(\vec{p},\beta))
   (\phi_x - \varphi_x)  \nonumber\\
&&  - e^{i\vec{p}\cdot\vec{x}}sh(g(\vec{p},\beta))(\tilde{\phi}_x - \varphi_x)]
+ie^{-i\vec{p}\cdot\vec{x}}ch(g(\vec{p},\beta))\Pi_x
-ie^{i\vec{p}\cdot\vec{x}}sh(g(\vec{p},\beta))\tilde{\Pi}_x \}
\end{eqnarray}
and
\begin{eqnarray}
A^\dagger_f(\vec{p},\beta) &=& ({\frac {1}{2(2\pi)^D f(\vec{p})}})^{1/2}\int_x
   \{f(\vec{p})[ e^{i\vec{p}\cdot\vec{x}}ch(g(\vec{p},\beta))
   (\phi_x - \varphi_x)  \nonumber\\
&& - e^{-i\vec{p}\cdot\vec{x}}sh(g(\vec{p},\beta))(\tilde{\phi}_x - \varphi_x)]
-ie^{i\vec{p}\cdot\vec{x}}ch(g(\vec{p},\beta))\Pi_x
+ie^{-i\vec{p}\cdot\vec{x}}sh(g(\vec{p},\beta))\tilde{\Pi}_x \} \;.
\end{eqnarray}
with the relations $[A_f(\vec{p},\beta),A^\dagger_f(\vec{p}',\beta)]
  =\delta (\vec{p}'-\vec{p})$ and $A_f(\vec{p},\beta)|\varphi,\beta>=0$. The
thermo-particle creation and annihilation operators of the tilde system can be
obtained from Eqs.(14) and (15) according to the rule of the tilde operation.

Letting the thermo-particle creation operator $A^\dagger_f(\vec{p},\beta)$ act
on the thermal vacuum Eq.(11) or (13), we can construct the one- and
two-thermo-particle excited states ( at the frame of the mass center ) as
\begin{equation}
|1> = A^\dagger_f (\vec{p},\beta)|\varphi,\beta>
\end{equation}
and
\begin{equation}
|2>
= \int d^D\vec{p}\Sigma(\vec{p})A^{\dagger}_f(\vec{p},\beta)
A^{\dagger}_f (-\vec{p},\beta)|\varphi,\beta> \;,
\end{equation}
respectively,
where $\Sigma(\vec{p})$ is the Fourier transform of the S-wave function of the
two thermo-particles with $\Sigma(-\vec{p})=\Sigma(\vec{p})$. Of course, one
can also manufacture the multi-thermo-particle states and the excited states of
the thermal tilde system. Here, we discuss them no more.

Next we are ready to calculate the FTGEP. From Ref.~\cite{8}, the entropy
operator is
\begin{equation}
S=\int dp [A_f(\vec{p})A^\dagger _f(\vec{p})ln(ch^2(g(\vec{p},\beta)))
          - A^\dagger_f(\vec{p})A_f(\vec{p})ln(sh^2(g(\vec{p},\beta)))]  \;,
\end{equation}
and hence, the Helmholtz free energy operator reads
\begin{equation}
F=\int_x {\cal N}_Q [{\cal H}_x]-{\frac {1}{\beta}} S \;.
\end{equation}
( Actually $k_B S$ is the entropy).
Using functional integration technique and the integral formula
$\int^\infty_0{\frac {2a}{\sqrt{\pi}}}e^{-a^2x^2}dx=1$, we obtain the average
value
\begin{equation}
<\varphi,\beta|{\cal N}_Q[V(\phi_x)]|\varphi,\beta>=
       \int^\infty_{-\infty}  {\frac{d\alpha}{2\sqrt{\pi}}}e^{-{\frac
    {\alpha^2}{4}}} V({\frac {\alpha}{2}}\sqrt{J_1(g)-I_1 (Q^2)}+\varphi_x)
\end{equation}
with the notation $J_n(g)=\int {\frac {dp f(\vec{p})ch(2g(\vec{p},\beta))}
{(2 \pi)^D (f(\vec{p}))^{2 n}}} \;.$ Thus making some functional integrations
leads to the vacuum average value of the free energy
\begin{eqnarray}
{\cal F}[\varphi,g,f; \beta]&=&<\varphi,\beta|F|\varphi,\beta>
     \nonumber \\
  &=&{\frac {1}{4}}J_0(g)+{\frac {1}{2}}(\nabla\varphi_x)^2
      +{\frac {1}{4}}\int
      {\frac {d^D\vec{p}}{(2\pi)^D}}{\frac {p^2}{f(\vec{p})}}
      ch(2 g(\vec{p},\beta))
  -{\frac {1}{2}}I_0 (Q^2)+{\frac {1}{4}}Q^2 I_1 (Q^2)
      \nonumber \\
&&+\int^\infty_{-\infty} {\frac {d\alpha}{2\sqrt{\pi}}}e^{-{\frac
     {\alpha^2}{4}}} V({\frac {\alpha}{2}}\sqrt{J_1(g)-I_1 (Q^2)}+\varphi_x)
     \nonumber \\
  &&   -{\frac {1}{\beta}}\int {\frac {dp}{(2 \pi)^D}}
     [ch^2 (g(\vec{p},\beta))ln(ch^2 (g(\vec{p},\beta)))
       - sh^2 (g(\vec{p},\beta))ln(sh^2 (g(\vec{p},\beta)))] \;\; ,
\end{eqnarray}
which is a functional of the functions $\varphi_x,g,f$. When the field system
Eq.(1) has a thermal equilibrium with the heat reservoir at $T\not=0$, its free
energy is absolutely minimized. So minimizing the free energy with respect to
$f(\vec{p})$ ( the Fourier transform of function $f_{xy}$ ) and taking
$\varphi_x=const.=\varphi$, we obtain
\begin{equation}
f(\vec{p})=f(p,\beta)
  =\sqrt{p^2+\mu^2(\varphi,\beta)}  \;,
\end{equation}
where
\begin{equation}
\mu^2(\varphi,\beta)
  = \int^\infty_{-\infty} {\frac {d\alpha}{2\sqrt{\pi}}}e^{-{\frac
 {\alpha^2}{4}}} V^{(2)}({\frac {\alpha}{2}}\sqrt{J_1(g)- I_1 (Q^2)}+\varphi)
\end{equation}
with $V^{(n)}(z)={\frac {d^nV(z)}{dz^n}}=
      \int^\infty_{-\infty}{\frac {d\Omega}{\sqrt{2\pi}}}(i\Omega)^n
          {\tilde{V}}(\Omega)e^{i\Omega z} \;.$
Thus, we see that the variational procedure has enforced the function $f$
depend both upon $\varphi$ and upon $\beta$ through the function $g$.

Furthermore, regardless of the relation Eq.(22) and independent of the last
variational procedure, one can minimize the free energy Eq.(21) with respect to
$g(\vec{p},\beta)$ and  obtain
\begin{equation}
g(\vec{p},\beta)=g(p,\beta)=ln\bigl({\frac {e^{{\frac {1}{2}}\beta
f(p,\beta)}+1}
{e^{{\frac {1}{2}}\beta f(p,\beta)}-1}}\bigr) \;.
\end{equation}
( Here, in the minimized result, we have made $f(p,\beta)$ take the place of
$f(\vec{p})$). Now the function $g$ is evidently related to $\beta$,
owing to the variational procedure. This point is similar to Ref.~\cite{11},
but different from Ref.~\cite{10}, where the expression between $\theta$ and
$\beta$ is artificially given ( $\theta$ there corresponds to $g$ here ) .
Substituting Eqs.(22) and (24) into Eq.(21), one has the FTGEP of Eq.(1)
\begin{eqnarray}
{\cal V}_T (\mu,\beta,\varphi) &\equiv& {\cal F}(\mu,\beta,\varphi)
={\cal F}[\varphi=const.,
g_{xy}(\beta),f_{xy}; \beta]\big|_{\mu^2(\varphi,\beta)\to\mu^2}
\nonumber  \\
&=& {\frac {1}{2}}J_0(g)-{\frac {\mu^2}{4}}J_1 (g)
  -{\frac {1}{2}}I_0 (Q^2)+{\frac {1}{4}}Q^2 I_1 (Q^2)
      \nonumber \\
&&+\int^\infty_{-\infty} {\frac {d\alpha}{2\sqrt{\pi}}}e^{-{\frac
     {\alpha^2}{4}}} V({\frac {\alpha}{2}}\sqrt{J_1(g)-I_1 (Q^2)}+\varphi)
     \nonumber \\
  &&-{\frac {1}{\beta}}\int {\frac {dp}{(2 \pi)^D}}
     [ch^2 (g(\vec{p},\beta))ln(ch^2 (g(\vec{p},\beta)))
    - sh^2 (g(\vec{p},\beta))ln(sh^2 (g(\vec{p},\beta)))] \;.
\end{eqnarray}
In this equation, ``$\to$'' means that $\mu$ takes the place of
$\mu(\varphi,\beta)$. Because of the nature of the minimizing procedure, $\mu$
should be chosen from the non-zero root of Eq.(23) and two
end points of the range $0<\mu<\infty$ so that ${\cal V}_T [\mu,\beta,\varphi]$
is an absolute minimum. Besides, sometimes the non-zero solution of Eq.(23) is
multi-valued, and so in that case, the suitable root should be decided
according to the stability condition
\begin{eqnarray}
&&{\frac {\partial^2 {\cal F}[\varphi,g,f; \beta]}{(\partial\mu^2)^2}} \nonumber \\
&&={\frac {1}{8}}J_2 (g) [1 + {\frac {1}{8}} J_2 (g)
 \int^\infty_{-\infty} {\frac {d\alpha}{2\sqrt{\pi}}}e^{-{\frac
 {\alpha^2}{4}}} V^{(4)}({\frac {\alpha}{2}}\sqrt{J_1(g)-
 I_1(Q^2)}+\varphi)] >0  \;.
\end{eqnarray}
For terms comprising $J_n(g)$'s and $I_n (Q^2)$'s in Eqs.(23), (25) and (26),
we can write them as
\begin{eqnarray}
&&{\frac {1}{2}}J_0(g)-{\frac {1}{4}}\mu^2 J_1(g)
     -{\frac {1}{2}}I_0 (Q^2)+{\frac {1}{4}}Q^2 I_1 (Q^2) \nonumber \\
 =&& {\frac {1}{2}}[I_0(\mu^2)-I_0(Q^2)] -{\frac {1}{4}}\mu^2 I_1(\mu^2)+
      {\frac {1}{4}}Q^2I_1(Q^2) + C_0-{\frac {\mu^2}{2}}C_2
      \nonumber \;,
\end{eqnarray}
 $$J_1(g)-I_1(Q^2)=I_1(\mu^2)-I_1(Q^2)+2 C_1 \;,$$
 and
 \begin{eqnarray}
&& {\frac {1}{\beta}}\int {\frac {dp}{(2 \pi)^D}}
     [ch^2 (g(p,\beta))ln(ch^2 (g(p,\beta)))
          - sh^2 (g(p,\beta))ln(sh^2 (g(p,\beta)))] \nonumber \\
&&  =C_0-{\frac {1}{\beta}}\int {\frac {d^D\vec{p}}{(2\pi)^D}}
  ln(1-e^{-\beta f(p,\beta)}) \;\;,\nonumber
\end{eqnarray}
where $C_n\equiv \int {\frac {d^D\vec{p}}{(2\pi)^D}}
{\frac {f^{1-n}(p,\beta)}{e^{\beta f(p,\beta)}-1}} $. Obviously, it is perhaps
hard for $C_0, C_1$ and $C_2$ to have simple analytic expressions, but they
are all finite. Moreover, for the case of $D<3$, although $I_n(\mu^2(\beta,
\varphi))$'s and $I_n (Q^2)$'s are divergent, the divergences can all cancelled
each other in the above three expressions. So when $D<3$, Eqs.(23),(25) and
(26) have not any divergence, and accordingly no further renormalization
procedures are needed. This indicates that the finite temperature field theory
with Eq.(1) is renormalizable for the case of $D<3$, at least within the
framework of the FTGWFA.

Now we are in a position for calculating one- and two-thermo-particle
energies. For thermo-particle excited states, thermofield dynamics uses the
following Hamiltonian \^{H} of the combined system of the $T=0$ field system
and its tilde partner
\begin{equation}
{\cal N}_Q [\hat{H}]\equiv \int_x \{{\cal N}_Q[{\cal H}_x]-
          {\cal N}_Q[\tilde{{\cal H}}_x]\} \;,
\end{equation}
and each thermo-particle excited state is the eigenstate of
${\cal N}_Q[\hat{H}]$. Similar to the calculation of the FTGEP, using
functional integration technique, one can obtain the one-thermo-particle
energy
\begin{equation}
m_1(\varphi,\beta)={\frac {<1|{\cal N}_Q[\hat{H}]|1>}{<1|1>}}
=\sqrt{p^2+\mu^2(\varphi,\beta)}
\end{equation}
and the two-thermo-particle energy
\begin{eqnarray}
m_2&\equiv&{\frac {<2|{\cal N}_Q[\hat{H}]|2>}{<2|2>}}
\nonumber \\
   &=&{\frac {2\int d^D\vec{p}\Sigma^2(p,\beta)f(p,\beta)
    +{\frac {\nu^{(4)}(\varphi,\beta)}{8(2\pi)^D}}
   \int{\frac {d^D\vec{p}\Sigma(p,\beta)}{f(p,\beta)}}
   \int{\frac {d^D\vec{p}\Sigma(p,\beta) ch(2g(p,\beta))}{f(p,\beta)}}}
    {\int d^D\vec{p}[\Sigma(p,\beta)]^2}}\;,
\end{eqnarray}
where
\begin{equation}
\nu^{(n)}(z,\beta)=\int^\infty_{-\infty} {\frac {d\alpha}{2\sqrt{\pi}}}
     e^{-{\frac {\alpha^2}{4}}} V^{(n)}({\frac {\alpha}{2}}\sqrt{J_1(g)
           -I_1(Q^2)}+z) \;.
\end{equation}
Minimizing $m_2$ with respect to $\Sigma(p,\beta)$ leads to a second kind of
Fredholm integral equation about $\Sigma(p,\beta)$ \cite{15}, and solving it,
we have
\begin{equation}
\Sigma(p,\beta)={\frac {C}{f(p,\beta)(2f(p,\beta)-m_2)}}
\biggl[ch(2g(p,\beta))+{\frac {\int
     {\frac {dp ch^2(2g(p,\beta))}{f^2(p,\beta)(2f(p,\beta)-m_2)}}}
{1+{\frac {\nu^{4}(\varphi,\beta)}{16(2\pi)^D}}
\int {\frac {dp ch(2g(p,\beta))}{f^2(p,\beta)(2f(p,\beta)-m_2)}}}}\biggr]
\end{equation}
with $C$ some normalization constant. Substituting the last expression into
Eq.(29) and noting the normalization of C, one can compute the energy $m_2$.
From Eq.(28), one can see that $\mu(\varphi,\beta)$ is just the mass of one
thermal particle. Thus, Eq.(23) can be used for calculating the mass of a
single thermo-particle. Moreover, the two terms in Eq.(29) can be interpreted
as the kinetic energy of the two constituent thermal particles and their
interacting energy, respectively. One can perhaps dicuss thermo-particle bound
states and scattering states with the help of Eq.(29) \cite{3} (1993).

In this letter, we have developed the GWFA of thermofield dynamics in $(D+1)$
dimensions, and calculated the FTGEP, one- and two-thermo-particle-state
energies of the system Eq.(1). Equations (25),(28) and (29) with some relevant
relations are convenient for an investigation of any concrete model involved
into the class of Eq.(1) ( including some bosonized models in condensed
matter physics \cite{16} ). When Eq.(25) is applied to $\lambda\phi^4$ and
$\phi^6$ models, one can obtain the same, corresponding results as in
Ref.~\cite{7} ( Roditi, Okopinska, Hajj and Cea ). Again, contrasting the
results in Ref.~\cite{17} with ones here, and noting the above-mentioned
analysis about $J_n (g)$'s and $I_n (Q^2)$'s, one can have the replacement
rule between the FTGEP and GEP which indicated in Ref.~\cite{7} ( $J_n(g)\to
I_n (\mu^2(\varphi))$). Still, the $T=0$ limit of each of Eqs.(25), (28) and
(29) is identical with that in Ref.~\cite{17}. Moreover, employing Eqs.(5),
(6), (9) and (10), one can find that the exponential factor in Eq.(11) is just
field-operator form of a thermal Bogoliubov transformation, and therefore
conclude that the TDGWFA is equivalent to the TDBTT. In fact, when $I_n (Q^2)'s$ are deleted, Eq.(25) can give the
non-renormalized equation Eq.(33) of Ref.~\cite{11}. Besides, if a non-uniform
background field is considered, the term ${\frac {1}{2}}(\nabla\varphi_x)^2$
will be added to Eq.(25) with $\varphi_x$ replacing $\varphi$, and then one can
have the following equation by ${\frac {\partial ({\cal V}_T [\mu,\beta,
\varphi_x])}{\partial \varphi_x}} =0$
\begin{equation}
\nabla^2\varphi_x- \int^\infty_{-\infty} {\frac {d\alpha}{2\sqrt{\pi}}}
     e^{-{\frac {\alpha^2}{4}}} V^{(1)}({\frac {\alpha}{2}}
     \sqrt{J_1(g)-I_1 (Q^2)}+\varphi_x)=0  \;,
\end{equation}
the $T=0$ limit of which is identical with that of Ref.~\cite{17}. This
equation can be used for considering quantum thermal solitons. Basing on
thermal soliton solution, Eqs.(28) and (29) can perhaps give excited-state
results of thermo-solitons. Of course, when the non-uniform background field is
considered, the discussion about the difference between the BTT and GWFA in
Ref.~\cite{17} is valid at the finite temperature case. Finally, the GWFA of
thermofield dynamics can be generalized to the time-dependent case so
as to investigate dynamics of statistical mechanical systems.

\acknowledgments
This work was finished at the Abdus Salam International Centre for Theoretical
Physics (ICTP), and the author would like to thank the Abdus Salam ICTP for
hospitality. This project was supported by the President Foundation of
Shanghai Jiao Tong University.

\end{document}